\documentclass[twocolumn,tighten]{aastex61}

\shorttitle{Recovery schemes for GRMHD}
\shortauthors{Siegel et al.}

\usepackage{amsfonts,amsmath,amssymb,mathrsfs}
\usepackage{graphicx}
\usepackage{dcolumn}
\usepackage{bm}
\usepackage{hyperref}
\hypersetup{
  colorlinks=true,        
  linkcolor=blue,         
  citecolor=blue,         
}

\usepackage{natbib}

\newcommand{\nb}{n_\mathrm{b}}
\newcommand{\nel}{n_\mathrm{e}}

\begin{document}

\title{Recovery schemes for
  primitive variables in \\general-relativistic magnetohydrodynamics}

\author{Daniel M. Siegel}
\altaffiliation{NASA Einstein Fellow}
\affiliation{Department of Physics, Columbia University, New York, NY 10027, USA
}
\affiliation{Columbia Astrophysics Laboratory, Columbia University, New York, NY 10027, USA
} 
\author{Philipp M\"osta}
\altaffiliation{NASA Einstein Fellow}
\affiliation{Department of Astronomy, University of California at Berkeley, Berkeley, CA 94720, USA
}
\author{Dhruv Desai}
\affiliation{Department of Physics, Columbia University, New York, NY 10027, USA
}
\author{Samantha Wu}
\affiliation{Department of Astronomy, University of California at Berkeley, Berkeley, CA 94720, USA
}

\date{\today}

\begin{abstract}
General-relativistic magnetohydrodynamic (GRMHD) simulations are an important tool to study a variety of astrophysical systems such as neutron star mergers, core-collapse supernovae, and accretion onto compact objects. A conservative GRMHD scheme numerically evolves a set of conservation equations for 'conserved' quantities and requires the computation of certain primitive variables at every time step. This recovery procedure constitutes a core part of any conservative GRMHD scheme and it is closely tied to the equation of state (EOS) of the fluid. In the quest to include nuclear physics, weak interactions, and neutrino physics, state-of-the-art GRMHD simulations employ finite-temperature, composition-dependent EOSs. While different schemes have individually been proposed, the recovery problem still remains a major source of error, failure, and inefficiency in GRMHD simulations with advanced microphysics. The strengths and weaknesses of the different schemes when compared to each other remain unclear. Here we present the first systematic comparison of various recovery schemes used in different dynamical spacetime GRMHD codes for both analytic and tabulated microphysical EOSs. We assess the schemes in terms of (i) speed, (ii) accuracy, and (iii) robustness. We find large variations among the different schemes and that there is not a single ideal scheme. While the computationally most efficient schemes are less robust, the most robust schemes are computationally less efficient. More robust schemes may require an order of magnitude more calls to the EOS, which are computationally expensive. We propose an optimal strategy of an efficient three-dimensional Newton--Raphson scheme and a slower but more robust one-dimensional scheme as a fall-back.
\end{abstract}


\section{Introduction}

Magnetized plasmas in gravitating systems where general relativity
is important play a pivotal role in understanding a variety of phenomena
across astrophysics, including accreting black holes in (X-ray)
binaries and active galactic nuclei, short and long gamma-ray
bursts, core-collapse supernovae, and compact-object mergers involving
neutron stars. Various general-relativistic magnetohydrodynamic (GRMHD)
codes have been developed to study these systems from first principles, both in stationary
background spacetime (e.g., HARM \citep{Gammie2003};
\citealt{Komissarov2005}; \citealt{Anton2006}; ECHO \citep{DelZanna2007}; 
Athena++ \citep{White2016}) and in fully dynamical spacetime
(e.g., \citealt{Duez05MHD0}; WhiskyMHD \citep{Giacomazzo2007}; \citealt{Cerda-Duran2008}; \citealt{Anderson2008};
\citealt{Kiuchi2012b}; GRHydro \citep{Moesta2014a}; SpEC
\citep{Muhlberger2014}; IllinoisGRMHD \citep{Etienne2015a}).

Conservative GRMHD schemes, such as the ones employed by the
aforementioned codes, solve the equations of GRMHD as a set of
conservation equations of the form (see Sec.~\ref{sec:recovery_problem})
\begin{equation}
  \partial_t(\sqrt{\gamma}\mathbf{q})
  + \partial_i[\alpha\sqrt{\gamma}\mathbf{f}^{(i)}(\mathbf{p})] =
  \mathbf{s}(\mathbf{p}),
\end{equation}
where $\alpha$ and $\gamma$ are (known) quantities derived from the
spacetime metric. Here, $\mathbf{q}$ represents a vector of so-called
conserved quantities, which are analytic functions of the
so-called primitive variables $\mathbf{p}$,
$\mathbf{q}=\mathbf{q}(\mathbf{p})$. Since the fluxes
$\mathbf{f}^{(i)}(\mathbf{p})$ and source terms
$\mathbf{s}(\mathbf{p})$ are functions of the primitives, at every
time step of the conservative scheme, the given vector of conserved
variables needs to be inverted to obtain the primitive variables in
order to evolve $\mathbf{q}$ to the next
time step. This recovery procedure therefore represents a core part
of every conservative GRMHD scheme.

Whereas the inversion from conservative to primitive quantities is
known in terms of analytical relations in Newtonian MHD and only
requires iterative procedures to obtain thermodynamic quantities for a
general equation of state (EOS), there is no known inversion in closed form in GRMHD, and
usually a set of nonlinear equations needs to be solved to obtain the
primitives $\mathbf{p}=\mathbf{p}(\mathbf{q})$. Complication arises from the
EOS, which captures the thermodynamic properties of the
fluid and which is closely tied to the recovery problem. Most GRMHD codes have
used simple analytic EOSs so far, such as (piecewise) polytropic EOS or an ideal-gas EOS, in which case specific recovery schemes exist \citep{Noble2006}. Even
in the context of simple analytic and well-behaved EOSs, this recovery problem
is one of the most error-prone key parts of any GRMHD evolution code.

Advancing realism in numerical simulations,
particularly in fully dynamical spacetime
applications such as core-collapse supernovae and compact binary mergers,
requires GRMHD codes to support composition-dependent,
finite-temperature EOSs,
typically formulated in terms of rest-mass density, temperature, and
electron fraction (three-parameter EOSs). For example, the inclusion of
weak interactions and
(approximate) neutrino transport into GRMHD requires such
EOSs. Furthermore, capabilities for composition-dependent, finite-temperature EOSs are
required for exploring the influence of different
EOSs on neutron star mergers; this may provide an important tool for inferring
the unknown EOS of nuclear matter at high densities from observations of neutron star mergers by the Laser
Interferometer Gravitational-wave Observatory (LIGO) and
electromagnetic facilities. To date, only very few GRMHD codes and
studies using composition-dependent EOSs exist (e.g.,
\citealt{Neilsen2014,Palenzuela2015,Moesta2014b,Moesta2015,Siegel2017a,Siegel2018a}). 

Composition-dependent EOSs are usually supplied to the code partially or entirely in the form of
tables. EOS calls by the code therefore involve table lookups with
interpolation operations, which can be computationally
expensive. Widely used recovery schemes that, in
principle, support three-parameter EOSs, such as the 2D scheme in \citet{Noble2006}
or the recovery scheme in \citet{Anton2006}, require additional
inversions using the EOS at every iteration step of the scheme from, e.g., specific enthalpy or
pressure to temperature, and thus introduce many additional EOS table
lookups; this potentially makes the recovery process computationally
expensive and introduces many more operations and additional sources of
error and/or failure that make the GRMHD scheme less robust.

In this paper, we discuss and compare several recovery schemes that are suitable for
composition-dependent, finite-temperature EOSs. Most of the schemes presented here have
already been used in GRMHD simulations, in different codes and by
different groups. Here, we provide the first comparison of these
schemes in a well-defined test setting,
i.e., independent of a specific application, and assess these schemes
in terms of three cardinal criteria: (i) speed (in terms of the number of
iterations and the number of EOS calls needed to converge), (ii) accuracy, and
(iii) robustness. We note that if the numerical code employing the recovery schemes discussed here is based on a finite-volume discretization (as is the case for typical GRMHD codes), the recovery methods discussed here are second-order accurate; additional complexity would arise for higher-order finite-volume schemes.

This paper is organized as follows. Section~\ref{sec:recovery_problem}
defines the recovery problem and Sec.~\ref{sec:recovery_schemes}
introduces the recovery schemes considered in this analysis. Results of our
comparison are presented in Sec.~\ref{sec:comparison} and
Sec.~\ref{sec:conclusion} summarizes our conclusions.

\section{The recovery problem}
\label{sec:recovery_problem}

The equations of ideal GRMHD include energy and momentum conservation, baryon number conservation, lepton number conservation, and Maxwell's equations (e.g., \citealt{Anton2006}),
\begin{eqnarray}
\nabla_\mu T^{\mu\nu} &=& Q u^\nu,
                          \label{eq:ev1}\\
\nabla_\mu (\nb u^\mu) &=& 0,
                          \label{eq:ev2}\\
\nabla_\mu (\nel u^\mu) &=& R,
\label{eq:ev3}\\
\nabla_\nu F^{*\mu\nu} &=& 0
\label{eq:ev4}
\end{eqnarray}
where $T^{\mu\nu}$ is the energy--momentum tensor, $u^\mu$ the
4-velocity, $\nb$ the baryon number density, $\nel$ the electron
number density, and $F^{*\mu\nu}$ the dual of the Faraday electromagnetic
tensor. The energy--momentum tensor in ideal GRMHD can be written as
\begin{equation}
  T^{\mu\nu}= \left(\rho h + b^2\right) u^\mu u^\nu + \left(p +
  \frac{b^2}{2}\right) g^{\mu\nu} - b^\mu b^\nu, \label{eq:Tmunu}
\end{equation}
where $p$ denotes pressure, $h=1+\epsilon + p/\rho$ specific enthalpy,
with $\epsilon$ being the specific internal energy, and $b^\mu\equiv (4\pi)^{-\frac{1}{2}}F^{*\mu\nu}u_\nu$ the
magnetic field vector in the frame comoving with the fluid,
$b^2\equiv b^\mu b_\mu$, and $g_{\mu\nu}$ the space-time
metric.\footnote{In this paper, Greek indices take space-time values 0, 1, 2, 3, whereas
Roman indices represent the spatial components 1, 2, 3 only. Repeated
indices are summed over.} We assume that the thermodynamic properties
of matter can be
described by a composition-dependent, finite-temperature EOS, formulated as a
function of density $\rho = \nb m_\mathrm{b}$, where $m_\mathrm{b}$
denotes the baryon mass, temperature $T$, and electron fraction
$Y_e=\nel \nb^{-1}$. Accordingly, we have added an evolution equation for
the electron fraction (Eq.~(\ref{eq:ev3})) to
the original set of ideal GRMHD evolution equations (\ref{eq:ev1}),
(\ref{eq:ev2}), and (\ref{eq:ev4}). The
terms on the right of Eqs.~(\ref{eq:ev1}) and (\ref{eq:ev3})
represent source terms that reflect the evolution of the electron
fraction due to weak interactions, which lead to the emission
of neutrinos and antineutrinos that carry away energy and momentum
from the system.

For numerical evolution, we adopt a 3+1 split of spacetime into non-intersecting
spacelike hypersurfaces of constant coordinate time $t$
\citep{Lichnerowicz44,Arnowitt2008}, writing the line element as
\begin{equation}
  \mathrm{d}s^2 = -\alpha^2 \mathrm{d}t^2 + \gamma_{ij}(\mathrm{d}x^i
  + \beta^i\mathrm{d}t) (\mathrm{d}x^j
  + \beta^j\mathrm{d}t),
\end{equation}
where $\alpha$ denotes the lapse function, $\beta^i$ the shift vector,
and $\gamma_{ij}$ the metric induced on every spatial
hypersurface. The hypersurfaces are characterized by the timelike unit normal
$n^\mu$, where $n^\mu = (\alpha^{-1},-\alpha^{-1}\beta^i)$ and $n_\mu =
(-\alpha,0,0,0)$. We adopt an Eulerian formulation of the equations of
GRMHD in terms of the Eulerian observer, defined as
the observer in the 3+1 foliation of spacetime moving with
4-velocity $n^\mu$ perpendicular to the hypersurfaces of constant
coordinate time $t$.

In this 3+1 decomposition of spacetime and adopting the Eulerian observer,
Eqs.~(\ref{eq:ev1})--(\ref{eq:ev4}) can be reformulated as a set of
conservation equations of the form
\begin{equation}
  \partial_t(\sqrt{\gamma}\mathbf{q})
  + \partial_i[\alpha\sqrt{\gamma}\mathbf{f}^{(i)}(\mathbf{p},\mathbf{q})] =
  \mathbf{s}(\mathbf{p}), \label{eq:GRMHDeqns}
\end{equation}
suitable for numerical evolution. Here, $\gamma$ is the determinant of the spatial metric $\gamma_{ij}$ and
\begin{equation}
  \mathbf{q} \equiv [D,S_i,\tau,B^i,DY_e] \label{eq:q}
\end{equation}
denotes the vector of conserved variables, where
\begin{eqnarray}
  D &\equiv& \rho W \label{eq:D}\\
  S_i &\equiv& -n_\mu T^{\mu}_{\phantom{\mu}i} = \alpha
               T^{0}_{\phantom{0}i}= (\rho h + b^2) W^2 v_i - \alpha b^0 b_i \label{eq:Si}\\
 \tau &\equiv& n_\mu n_\nu T^{\mu\nu} - D = \alpha^2 T^{00}-D\\
  &=& (\rho h + b^2)W^2 - \left(p +\frac{b^2}{2}\right) -
         \alpha^2(b^0)^2 - D, \label{eq:tau}
\end{eqnarray}
and $B^i$ are the 3-vector components of the magnetic field $B^\mu
\equiv (4\pi)^{-\frac{1}{2}}F^{*\mu\nu}n_\nu$ as measured by the
Eulerian observer; $D$ is called the conserved density, $S_i$
represent the conserved momenta, and $\tau$ is the
conserved energy. For further reference, we note that
\begin{eqnarray}
  S^2 &\equiv& S^i S_i \nonumber\\
  &=& (\rho h + b^2)^2 W^4 v^2 + \alpha^2 (b^0)^2\nonumber\\
   &&\times\left[b^2(1-2W^2) -
      2\rho h W^2 + \alpha^2(b^0)^2\right]. \label{eq:S2}
\end{eqnarray}
The Eulerian observer measures the fluid
4-velocity $u^\mu = (u^0,u^i)$ with corresponding 3-velocity
\begin{equation}
  v^i \equiv \frac{\gamma^i_{\phantom{i}\mu}u^\mu}{-u^\mu n_\mu} =
  \frac{u^i}{W} + \frac{\beta^i}{\alpha}, \mskip20mu  v_i = \frac{\gamma_{i\mu}u^\mu}{-u^\mu n_\mu} =
  \frac{u_i}{W}, \label{eq:vi}
\end{equation}
where
\begin{equation}
  W \equiv -u^\mu n_\mu = \alpha u^0 = \frac{1}{\sqrt{1-v^2}} \label{eq:W}
\end{equation}
is the relative Lorentz factor between $u^\mu$ and $n^\mu$, with
$v^2\equiv\gamma_{ij}v^iv^j$. We note that the comoving and Eulerian
magnetic field components are related by
\begin{eqnarray}
  b^i &=& \frac{B^i}{W} + b^0 (\alpha v^{i} - \beta^i), \\
  b_i &=& \frac{B_i}{W} + \alpha b^0 v_{i}, \\
  b^0 &=& \frac{W}{\alpha} B^iv_i, \label{eq:bi}
\end{eqnarray}
and
\begin{equation}
  b^2 = b^\mu b_\mu = \frac{B^2 + (\alpha b^0)^2}{W^2}, \label{eq:b2}
\end{equation}
where $B^2\equiv B^iB_i$. Furthermore, we shall refer to
\begin{equation}
  \mathbf{p} \equiv [\rho,v^i,\epsilon,B^i,Y_e] \label{eq:p}
\end{equation}
as the vector of primitive variables.

Conservative numerical GRMHD schemes evolve the state vector
$\mathbf{q}$ from one time step to the next using
Eq.~(\ref{eq:GRMHDeqns}). This involves computing the flux terms
$\mathbf{f}^{(i)}(\mathbf{p},\mathbf{q})$ and source terms
$\mathbf{s}(\mathbf{p})$ for given $\mathbf{q}$, for which one needs to obtain the primitive
variables $\mathbf{p}$ from the conserved ones. While the conservative
variables as a function of primitive variables, $\mathbf{q}
=\mathbf{q}(\mathbf{p})$, are given in analytic form by
Eqs.~(\ref{eq:D})--(\ref{eq:b2}), the inverse relation $\mathbf{p}
=\mathbf{p}(\mathbf{q})$, i.e., the recovery of primitive variables
from conservative ones, is not known in closed form; this rather
requires numerical inversion of the aforementioned set of
nonlinear equations.

\section{Recovery schemes}
\label{sec:recovery_schemes}

The recovery problem is, in principle,
five-dimensional (5D), i.e., obtaining the primitives
$\rho$, $v^i$, $\epsilon$ in Eq.~(\ref{eq:p}) from Eqs.~(\ref{eq:D})--(\ref{eq:b2})
involves numerical root-finding in a 5D space, given that the recovery
of the magnetic field components and the electron fraction from
$\mathbf{q}$ (see Eq.~(\ref{eq:q})) is trivial. While 5D schemes as
originally implemented in, e.g., the HARM code \citep{Gammie2003}, were
later shown to be slow and inaccurate, reducing the dimensionality of
the problem to 1D or 2D with the help of certain scalar quantities
proved more efficient, with the 2D schemes being less pathological and
therefore being recommended \citep{Noble2006}.

In this section, we first discuss the widely used 2D scheme by
\citet{Noble2006} and extend it to three-parameter EOSs, which will serve as
a standard reference (hereafter referred to as ``2D NR Noble''). We then
introduce a variant of the 2D scheme used in
\citet{Anton2006}, \citet{Aasi2014}, \citet{Giacomazzo2007}, and \citet{Cerda-Duran2008}, suitable for the use
of three-parameter EOSs (henceforth ``2D NR''). Furthermore, we discuss
several recovery schemes (3D and effective 1D) that have been
proposed recently and that are
used in current GRMHD codes with support for composition-dependent, finite-temperature EOSs. 

\subsection{2D NR Noble et al. scheme}
\label{sec:2D_Noble}

This scheme reduces the dimensionality of the recovery problem by
making use of certain scalar quantities that can be computed from the
conservatives; it is based on Eqs.~(27) and (29) of \citet{Noble2006}, which are two equations in the unknowns $z\equiv \rho h W^2$ and $v^2$. In our notation, they read
\begin{equation}
  v^2(B^2 + z)^2 - \frac{(B^iS_i)^2(B^2 + 2z)}{z^2}
  - S^2 = 0, \label{eq:2DNoble_1}
\end{equation}
\begin{multline}  
  \tau + D - \frac{B^2}{2}(1+v^2) + \frac{(B^iS_i)^2}{2z^2}\\
   - z + p(z,v^2) = 0. \label{eq:2DNoble_2}
\end{multline}
The pressure $p(z,v^2)$ can be directly obtained in terms of the unknowns
$z$ and $v^2$ only for simple analytic EOSs
such as the ideal-gas EOS. In order to extend this scheme to
three-parameter EOSs, at every iteration of the scheme we invert the specific enthalpy $h = z/(\rho W^2)$ for given $\rho = D/W$, $W=(1-v^2)^{-1/2}$, and $Y_e$ with the help of the EOS to find the temperature $T$ and then compute $p(\rho,T,Y_e)$ using the EOS. This involves inversion with a 1D Newton--Raphson (NR) scheme, which falls back to bisection if the inversion fails.

Once converged, the final primitives $\rho$,
$v^i$, and $\epsilon$ can be obtained from $\rho = D\sqrt{1-v^2}$,
\begin{equation}
   v^i = \frac{\gamma^{ij}S_j}{z+B^2}  + \frac{(B^jS_j)
     B^i}{z(z+B^2)}, \label{eq:2D_vi}
\end{equation}
 and, using the EOS,
\begin{equation}
  \epsilon = \epsilon(\rho,T,Y_e). \label{eq:2D_eps}
\end{equation}

\subsection{2D NR scheme}
In the following, we present a 2D scheme that starts from the same equations
as the 2D schemes in \citet{Anton2006},\citet{Giacomazzo2007}, and \citet{Cerda-Duran2008}
and is similar to \citet{Noble2006},
but formulated in terms of unknowns that are more suitable
for the use of modern microphysical (tabulated) EOSs provided
in terms of $\rho$, $T$, and $Y_e$. The advantage of our formulation is that additional inversion steps from specific enthalpy or specific internal energy to
temperature are not necessary. Such
additional inversions at every
iteration of the main root-finding process are not required in the case of
the ideal-gas EOS, but introduce additional operations and
sources of error or failure for three-parameter EOSs, and are computationally expensive.

Similar to the 2D NR Noble scheme (Sec.~\ref{sec:2D_Noble}), this scheme reduces
the dimensionality of the recovery problem to 2D
by making use of the two scalar quantities $\tau$
(see Eq.~\eqref{eq:tau}) and $S^2$ (see Eq.~\eqref{eq:S2}). Setting $z\equiv \rho h W^2$ as before and using the identity
\begin{equation}
  B^i S_i = \rho h W \alpha b^0,
\end{equation}
one can write Eq.~(\ref{eq:tau}) and (\ref{eq:S2}) as
\begin{equation}
  \left[\tau + D - z - B^2 + \frac{(B^iS_i)^2}{2z^2} + p \right]W^2 - \frac{B^2}{2} = 0, \label{eq:2D_f1}
\end{equation}
\begin{multline}  
  \left[(z+B^2)^2 - S^2 - \frac{2z + B^2}{z^2}(B^i S_i)^2 \right]W^2 \\-
  (z+B^2)^2 = 0. \label{eq:2D_f2}
\end{multline}
Employing the EOS to compute the pressure from $(\rho,T,Y_e)$, we take
Eqs.~(\ref{eq:2D_f1}) and (\ref{eq:2D_f2}) as two equations in
the unknowns $W$ and $T$ by setting $z=z(W,T)$, $p=p(W,T)$, with
$\rho$ being calculated from Eq.~(\ref{eq:D}), and solve using a 2D
NR algorithm. Once converged, the final primitives $\rho,
v^i, \epsilon$ are obtained from Eqs.~(\ref{eq:D}), (\ref{eq:2D_vi}),
and (\ref{eq:2D_eps}).

\subsection{3D NR scheme}
Equations (\ref{eq:2D_f1}) and (\ref{eq:2D_f2}) can be extended
to a 3D system in the unknowns $W$, $z$, and $T$ by adding a
constraint on the specific internal energy \citep{Cerda-Duran2008},
\begin{equation}
  \left[\tau + D - z - B^2 + \frac{(B^iS_i)^2}{2z^2} + p
  \right]W^2 - \frac{B^2}{2} = 0 \label{eq:3D_f1}
\end{equation}
\begin{multline}
  \left[(z+B^2)^2 - S^2 - \frac{2z + B^2}{z^2}(B^i S_i)^2 \right]W^2 \\
  -(z+B^2)^2 = 0 \label{eq:3D_f2}
\end{multline}
\begin{equation}  
  \epsilon - \epsilon(\rho,T,Y_e) = 0. \label{eq:3D_f3}
\end{equation}
Here, $\epsilon$ is computed as
\begin{equation}
  \epsilon = \epsilon(W,z,T) = h - 1 - \frac{p}{\rho} = \frac{z-DW - p
  W^2}{DW}, \label{eq:3D_eps}
\end{equation}
and both $p$ in Eqs.~(\ref{eq:3D_f1}) and (\ref{eq:3D_eps}) and
$\epsilon(\rho,T,Y_e)$ in Eq.~(\ref{eq:3D_f3}) are computed using
the EOS with $\rho$ from Eq.~(\ref{eq:D}). As this scheme also employs the temperature directly as an unknown, it does not require any inversions with the EOS. Once the system of
Eqs.~(\ref{eq:3D_f1})--(\ref{eq:3D_f3}) has been solved with a 3D
NR scheme, one recovers the final primitives $\rho,
v^i, \epsilon$ from Eqs.~(\ref{eq:D}), (\ref{eq:2D_vi}), and (\ref{eq:2D_eps}).

\subsection{effective 1D schemes}

\subsubsection{Method of Palenzuela et al.}

The scheme recently presented by \citet{Neilsen2014} and \citet{Palenzuela2015}
reduces the dimensionality of the recovery problem to 1D, but requires
an additional inversion step using the EOS at
every iteration of the main scheme if the EOS is not provided in
terms of $(\rho,\epsilon,Y_e)$. The scheme utilizes Brent's method to
solve 
\begin{equation}
f(x) = x-\hat h\hat W=  x - \left( 1+\hat \epsilon + \frac{p(\hat \rho, \hat \epsilon, Y_e)}{\hat \rho} \right) \hat W
\label{eq:f_Brent}
\end{equation}
in the unknown
\begin{equation}
x \equiv \frac{\rho h W^2}{\rho W} =  hW,\\
\end{equation}
which is bounded by
\begin{equation}
1+q-s<x<2+2q-s.
\end{equation}
Here, following \citet{Palenzuela2015}, we have defined
\begin{eqnarray}
q &\equiv& \tau/D, \\
r &\equiv& S^2/D^2, \\
s &\equiv& B^2/D,   \\
t &\equiv& B_iS^i/D^{3/2}. \label{eq:defs_Brent}
\end{eqnarray}

Quantities with a hat in Eq.~(\ref{eq:f_Brent}) are computed at every
iteration step from $x$
and the conservatives by the following procedure:

\begin{itemize}
\item[(i)] With
\begin{equation}
  S_i = (\rho hW^2 + B^2) v_i -(B_jv^j)B_i
\end{equation}
(see Eqs.~(\ref{eq:Si}), (\ref{eq:bi}), (\ref{eq:b2})), using 
\begin{equation}
B^iS_i= \rho hW^2B^iv_i \label{eq:BiSi}
\end{equation}
and the definition of $r$
(see Eq.~(\ref{eq:defs_Brent})), one can write
\begin{equation}
  \hat W^{-2} = 1 - \frac{x^2 r + (2x+s)t^2}{x^2 (x+s)^2}. \label{eq:W_Brent}
\end{equation}

\item[(ii)] Using Eq.~(\ref{eq:D}), one obtains $\hat \rho = D / \hat W$.

\item[(iii)] Starting from $\epsilon = h - 1 - p/\rho$ using
  Eqs.~(\ref{eq:bi}), (\ref{eq:b2}), and (\ref{eq:BiSi}), one can
  write
\begin{eqnarray}
  \hat \epsilon &=& -1 + \frac{x}{\hat W}( 1-\hat W^2) \\
  &&+ \hat W \left[1+
    q-s+\frac{1}{2}\left(\frac{t^2}{x^2} +\frac{s}{\hat W^2}\right)
  \right]. \label{eq:eps_Brent}
\end{eqnarray}

\item[(iv)] Using the EOS, one inverts $\hat \epsilon$ to find the
  corresponding temperature and thus $p(\hat \rho, \hat \epsilon, Y_e)$.

\end{itemize}

Once converged, the recovered $\rho$ and $\epsilon$ are obtained from
Eqs.~(\ref{eq:D}), (\ref{eq:W_Brent}), and (\ref{eq:eps_Brent}); the
velocity components $v^i$ are obtained from Eq.~(\ref{eq:2D_vi}) setting
$z = x\rho W$.

\subsubsection{Method of Newman \& Hamlin}
\label{sec:Newman}

The recovery scheme presented in \citet{Newman2014} is an effective 1D method, as is the scheme of \citet{Palenzuela2015} and \citet{Neilsen2014} discussed above. It iterates over the fluid pressure to find a solution, and as in the method of \citet{Palenzuela2015} it requires an additional inversion step using the EOS for every
main scheme iteration. It is independent of an initial guess because it starts with $p=0$ as a guess for the pressure. For a tabulated EOS with a finite minimum pressure, we employ the pressure obtained from the minimum temperature of the EOS at given $\rho=D/W$ and $Y_e$, where $W$ is an initial guess for the present Lorentz factor.

\begin{itemize}
\item[(i)] Starting from the conserved quantities 
\begin{eqnarray}
	e &\equiv& \tau+D, \label{newman1}\\
	S^i &=& [wW^2+B^2]v^i - (B^i v_i) B^i,\label{newman2}\\
 D &=& \rho W, \label{newman3}\\
 \mathcal{M}^2 &=& \frac{(B^i v_i)^2}{\sqrt{\gamma}}, \mathcal{T} =\frac{ B^iS_i}{\sqrt{\gamma}}, B^2,\label{newman4}
\end{eqnarray}
where $w = \rho h$, the scheme solves a cubic polynomial  
\begin{equation}
	f(\varepsilon ) = \varepsilon^3 + a\varepsilon^2 + d
\label{newman5}
\end{equation}
for 	
\begin{equation}
\varepsilon \equiv B^2 + z, \\
\label{newman6}
\end{equation}
where $z\equiv\rho h W^2 = w W^2$ as before and
\begin{eqnarray}
a = e + P + \frac{B^2}{2}, \label{newman8}\\
d = \frac{1}{2}(\mathcal{M}^2 B^2 - \mathcal{T}^2). \label{newman9}
\end{eqnarray}
Evaluating the positive extremum of this equation, one can find a sufficient condition
for the existence of a positive root to Eq.~\eqref{newman5} as
\begin{equation}
	d \leq \frac{4}{27} 4 a^3.
\label{newman10}
\end{equation}
Using 
\begin{equation}
	d \equiv \frac{4}{27}a^3 \cos^2(\phi)
\label{newman11}
\end{equation}
one can show that for $ 0 \leq \phi \leq 2\pi$, with
\begin{equation}
\phi = \arccos\left(\frac{1}{a}\sqrt{\frac{27d}{4a}}\right),
\label{newman12}
\end{equation}
the first root of 
\begin{equation}
\varepsilon^{(l)} = \frac{1}{3}-  \frac{2}{3}a\cos\left(\frac{2}{3}\phi+ \frac{2}{3}l\pi\right)
\label{newman13}
\end{equation}
corresponds to the correct and physical solution of Eq.~\eqref{newman5}, from which one obtains $z$.

\item[(ii)] One can now compute primitive quantities as
\begin{eqnarray}
v^2 &=& \frac{\mathcal{M}^2z^2 + \mathcal{T}^2(B^2 + 2z)}{z^2 (B^2+z)^2}, \label{newman14} \\
W^2 &=& \frac{1}{\sqrt{1-v^2}}, \label{newman15}\\
w &=& \frac{z}{W^2}, \label{newman16}\\
\rho &=& \frac{D}{W}.\label{newman17}
\end{eqnarray}
Inverting $w$ with the help of the EOS for $\rho$ of Eq.~\eqref{newman17} and given $Y_e$, one obtains the corresponding temperature $T$ and a new pressure guess $p$.

\item[(iii)] One iterates over steps (i) and (ii) until the pressure has converged. When three successive pressure estimates are known, the Aitken acceleration scheme as described in \citet{Newman2014} is used. If convergence is reached under the Aitken acceleration, we recalculate the variables as in step (ii). If not, the solver restarts from the beginning of step (i).

\item[(iv)] Once a converged value for $p$ has been acquired, including a consistent set of $\rho$, $T$, and $\epsilon$ from (iii), the remaining primitives $v^i$ can be obtained from Eq.~\eqref{eq:2D_vi}.

\end{itemize}

\section{Comparison of recovery schemes}
\label{sec:comparison}

\subsection{Comparison setup and criteria}
\label{sec:comparison_setup}

Comparing the different recovery schemes described in
Sec.~\ref{sec:recovery_schemes} proceeds by generating a parameter
space of primitive variables $\mathbf{p} \equiv
[\rho,v^i,\epsilon,B^i,Y_e]$, computing the corresponding
conservative variables $\mathbf{q} = \mathbf{q}(\mathbf{p})$ using Eqs.~(\ref{eq:D})--(\ref{eq:tau}), and
applying the recovery schemes to obtain sets of primitives that can be
compared to the original ones. We perform this comparison of schemes
for three different EOSs---the analytic ideal-gas EOS $p = (\Gamma
-1)\rho\epsilon$ (where we arbitrarily set $\Gamma = 4/3$), the
partially analytic, partially tabulated Helmholtz EOS
\citep{Timmes1999,Timmes2000} with modifications as described in \citet{Siegel2017a}, and the fully tabulated Lattimer--Swesty
EOS \citep{Lattimer1991} with incompressibility modulus
$K=220\,\mathrm{MeV}$ (hereafter LS220), which serves as a widely used standard
reference EOS in GRMHD simulations of neutron star mergers and
core-collapse supernovae. We reduce the parameter space by varying the
Lorentz factor $W$ instead of $v^i$, and then choosing the orientation
of the 3-velocity randomly at every point in the parameter space. The
magnetic field vector is
set by a specified absolute value and oriented relative to the
3-velocity vector by a specified angle. For the results presented
below, we shall assume that $B^i$ and $v^i$ are aligned for
simplicity; the results for the comparison of the schemes do not
depend strongly on this relative orientation. For those recovery schemes that
require initial guesses explicitly (such as 2D NR Noble, 2D NR, 3D NR) or implicitly (such as 1D Brent), we apply a $5\%$
random perturbation to the initial primitives and use the perturbed
set as the initial guess; this $5\%$ deviation typically represents a conservative bracket on the change
from one time step to the next in an actual GRMHD evolution and should
thus lead to a conservative upper limit on, e.g., the number of
iterations or EOS calls required to converge to the solution. We have checked that the results presented here are essentially unchanged if an even higher perturbation of 10\% is applied. If the recovery fails in the case of 2D NR and 3D NR, we use the ``safe guess'' initial values of \citet{Cerda-Duran2008} in a second recovery attempt. We define convergence
of a scheme by having surpassed a $\mathrm{tol}_\mathrm{conv}
=5\times 10^{-9}$ (maximum) relative error in the
iteration variable(s). The resulting recovered primitives are compared
to the original set and to the results from the other recovery schemes
in terms of the following criteria.

\begin{itemize}
\item[(i)] \textit{Speed}. The speed of a given recovery scheme is assessed by
  counting the number of iterations needed to converge to a
  solution. Additionally, we have implemented counters to evaluate the total
  number of EOS calls required until convergence. As table lookups can
  dominate the computational cost of the recovery scheme and with some
  schemes requiring additional EOS inversion steps at each iteration,
  monitoring the number of EOS calls represents a better proxy for
  the overall speed of a scheme than just counting
  the number of iterations.

\item[(ii)] \textit{Accuracy}. We assess the accuracy of recovery by comparing
  the recovered primitives $\mathbf{p}$ with the original ones
  $\mathbf{p}_\mathrm{or}$ and computing the mean relative deviation/error,
  \begin{equation}
     \mathrm{tol}_\mathrm{recov}\equiv\frac{1}{5}\sum_{i=1}^5 \Big\vert
     \frac{\mathbf{p}_i-\mathbf{p}_{\mathrm{or},i}}{\mathbf{p}_{\mathrm{or},i}} \Big\vert,
  \end{equation}
  where only $\rho$, $v^i$, and $\epsilon$ are included in the summation, because
  $B^i$ and $Y_e$ are recovered trivially. We define successful
  recovery of the primitives as when all primitives have been recovered with a relative error smaller than
  $\mathrm{tol}_\mathrm{recov}=10 \times \mathrm{tol}_\mathrm{conv}$.

\item[(iii)] \textit{Robustness}. Robustness is assessed qualitatively
  in terms of the following criteria: independence of an initial
  guess, guaranteed convergence, and independence of thermodynamic
  derivatives.

\end{itemize}

\begin{figure*}[tb]
\centering
\includegraphics*[width=0.95\textwidth]{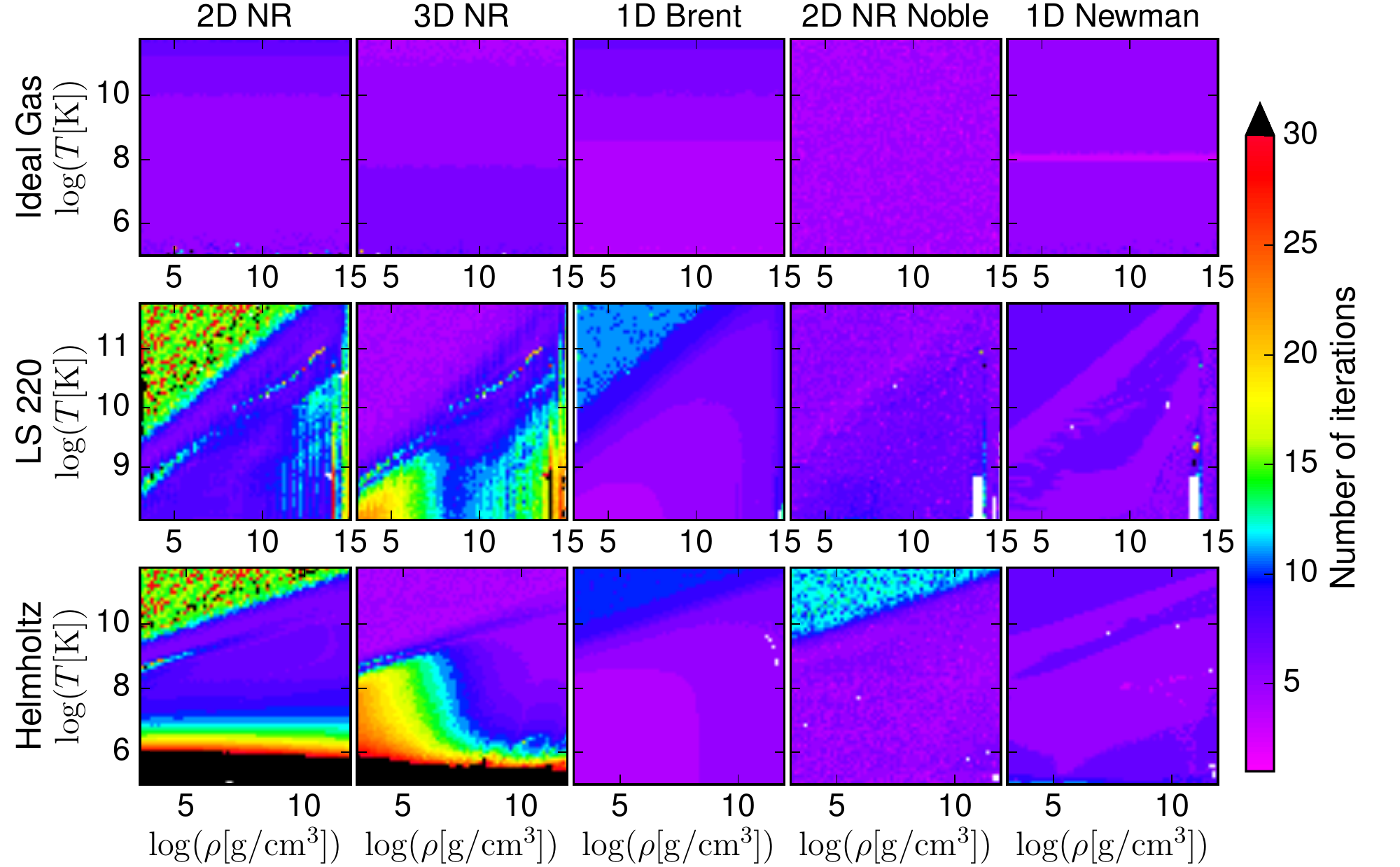}\\[0.1cm]
\includegraphics*[width=0.95\textwidth]{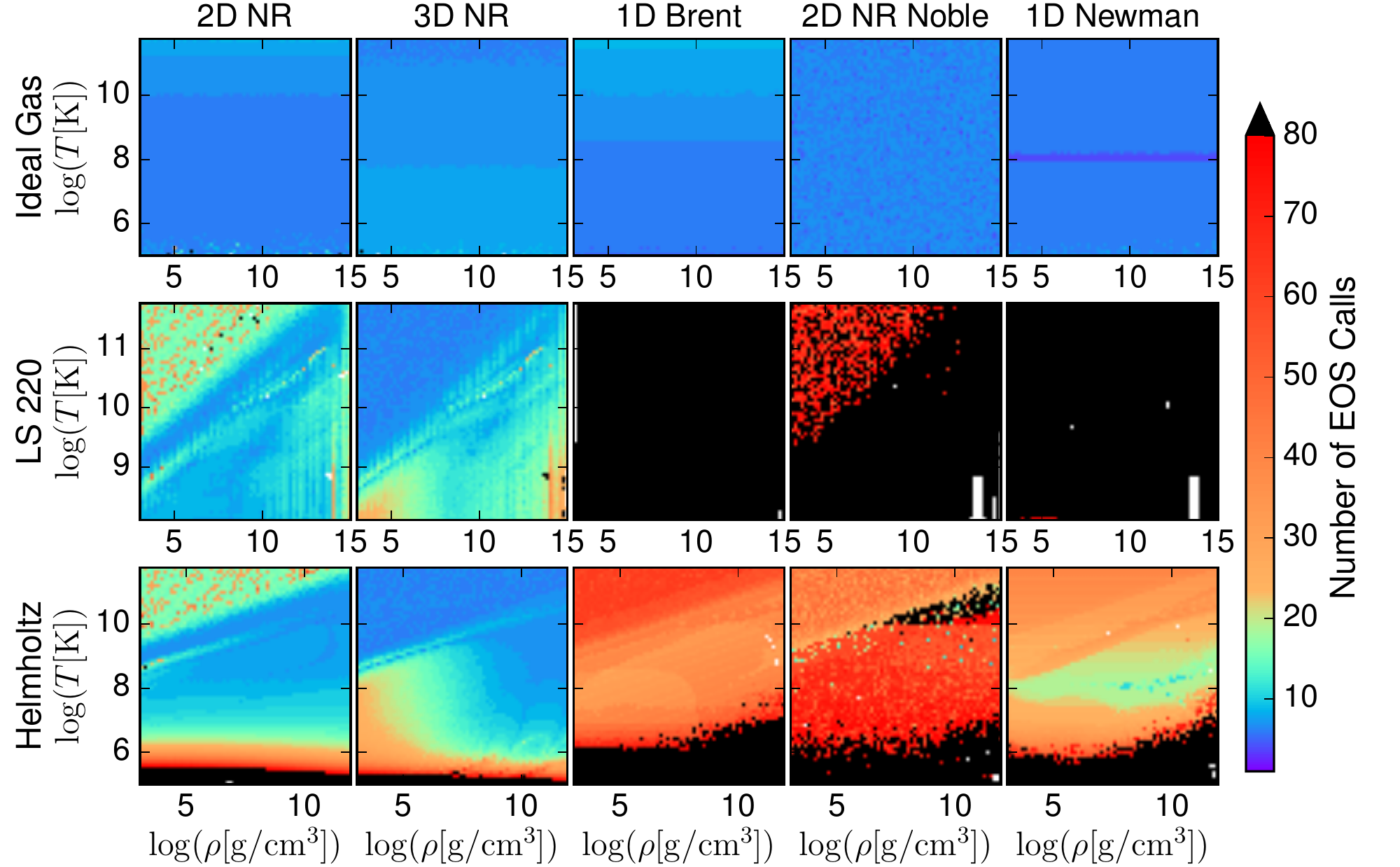}
\caption{Comparison of schemes in terms of the number of iterations (top) and number of EOS calls (bottom) required for convergence. This set of comparison assumes a Lorentz factor of $W=2$, magnetic to fluid pressure ratio of $p_\mathrm{mag}/p= 10^{-3}$, and an electron fraction of $Y_e=0.1$.}
\label{fig:W1-2_prat--3}
\end{figure*}

\begin{figure*}[tb]
\centering
\includegraphics*[width=0.95\textwidth]{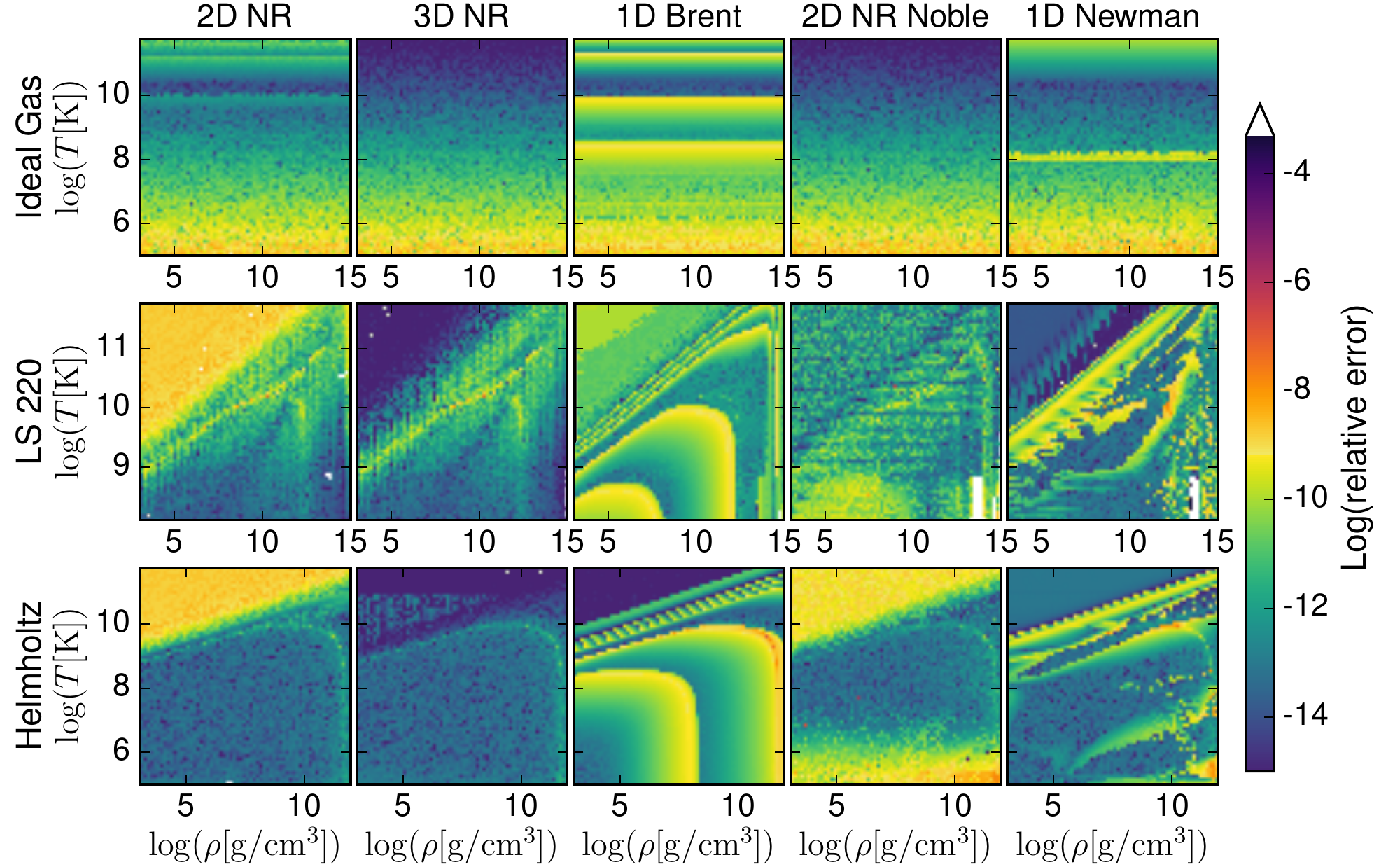}
\caption{Comparison of schemes in terms of the relative accuracy obtained in recovering the primitive variables, for the same set of parameters as in Fig.~\ref{fig:W1-2_prat--3}, i.e., $W=2$, $p_\mathrm{mag}/p= 10^{-3}$, $Y_e=0.1$.}
\label{fig:W1-2_prat--3_acc}
\end{figure*}

\subsection{Results}
\label{sec:results}
\vspace{0.5cm}

\begin{figure*}
\centering
\includegraphics*[width=0.95\textwidth]{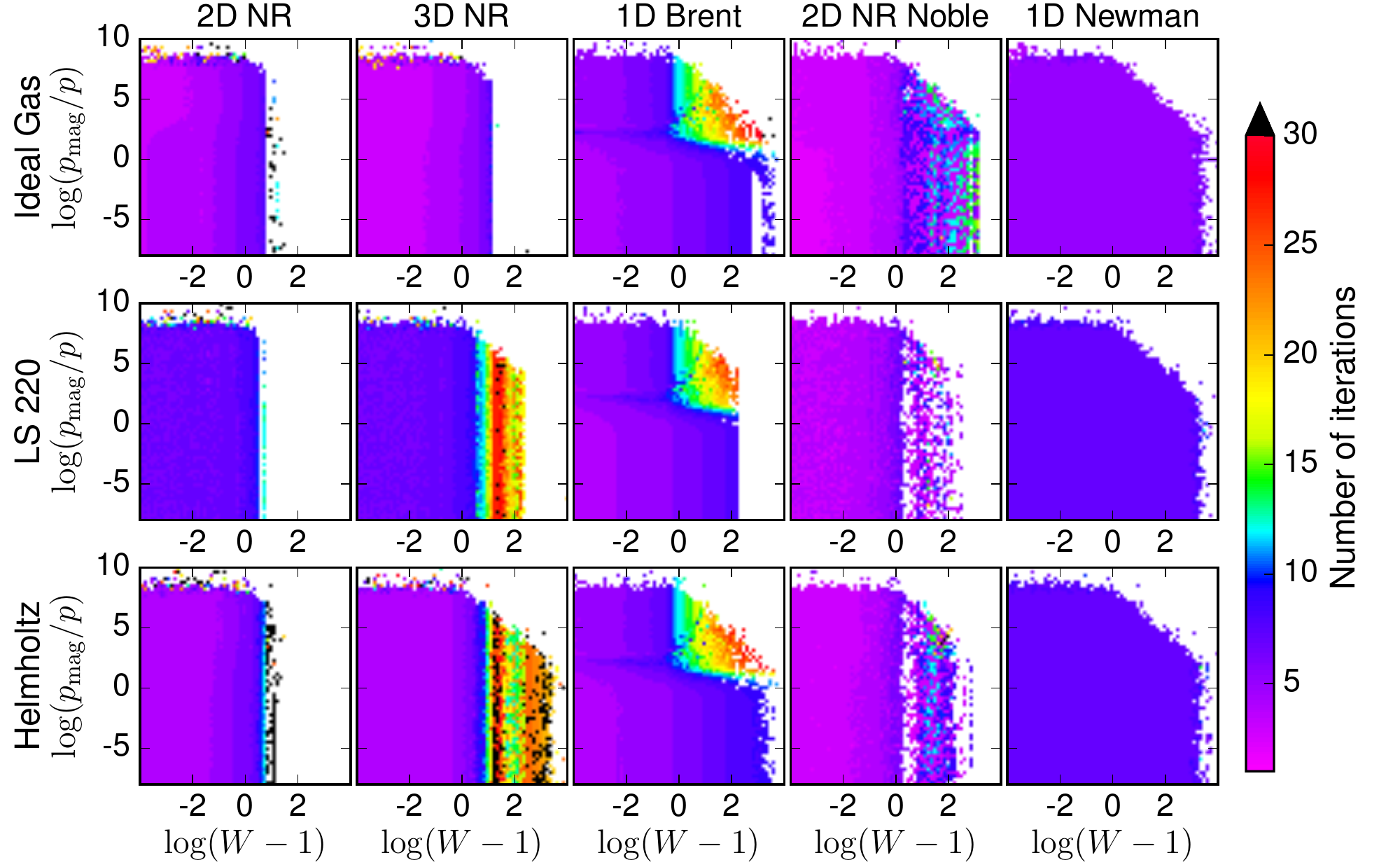}
\caption{Comparison of schemes in terms of the number of iterations required for convergence. This set of comparisons assumes a density of $\rho=10^{11}\,\mathrm{g}\,\mathrm{cm}^{-3}$, a temperature of $T=5\,\mathrm{MeV}\,(5.8\times 10^{10}\,\mathrm{K})$, and an electron fraction of $Y_e=0.1$.}
\label{fig:RHO-1e11_T-5_n_iter}
\end{figure*}

Figures~\ref{fig:W1-2_prat--3}--\ref{fig:RHO-1e11_T-5_n_iter} show 
comparisons of the five different recovery schemes discussed in this
paper. We present results for three different EOSs---the ideal-gas EOS, the
LS220 tabulated EOS, and the partially tabulated Helmholtz EOS. White
spaces correspond to parameter sets for which the recovery process either
failed or did not achieve the required accuracy $\mathrm{tol}_\mathrm{recov
}$ (Sec.~\ref{sec:comparison_setup}) within a certain number of iterations.

The top panels of Fig.~\ref{fig:W1-2_prat--3} compare the number of iterations required for each scheme to converge to the correct primitive variables within the required
accuracy $\mathrm{tol}_\mathrm{conv}$ (Sec.~\ref{sec:comparison_setup}). The parameter range in density $\rho$ and temperature $T$
is chosen to cover the validity ranges of the chosen EOS and we assume representative values of $W = 2$ and $p_\mathrm{mag}/p = 10^{-3}$, where $p_\mathrm{mag}=b^2/2$ is the magnetic pressure.

For the ideal-gas EOS, all schemes recover the entire parameter space in density and temperature with only a few iterations ($\sim 5$).
The 2D NR Noble scheme overall requires the smallest number of iterations.

For the LS220 EOS there is more variation among the different schemes.
The 2D NR scheme recovers most of the parameter space with $\lesssim\!10$ 
iterations but requires 15--30 iterations in the upper left corner as well as for
densities above $\sim\!10^{12}\,\mathrm{g\,cm^{-3}}$ and temperatures below
$\log T = 10.5\,\mathrm{K}$. The 3D NR scheme needs 15--25 
iterations for regions
at lower temperature, whereas the 1D Brent method performs well
and recovers the entire parameters space with $\lesssim\!10$ iterations. The
2D NR Noble and 1D Newman methods perform similarly with the exception
that there are small regions where the schemes do not recover the
primitives within the given accuracy at very high density and low 
temperature (white pixels in the bottom right corners). These points
account for less than 1.5\% of the total points. These failures arise 
because the additional inversion step of finding a temperature for the 
given relativistic enthalpy fails in the EOS table. We have investigated
this in some detail and find that for all of these points the initial 
first step in the solver leads to relativistic enthalpy values outside of 
the EOS table. We have implemented a limitation on step size for these cases 
but did not find a straightforward way to successfully apply it 
consistently to all points without additional side effects. While there 
are a number of ways in which a more sophisticated treatment of step sizes 
can be implemented, we have chosen to not fine-tune inversions for individual schemes too much with respect to others in order to ensure a fair 
comparison. The requirement of this additional inversion for temperature 
given the relativistic enthalpy is a feature/weakness of these schemes 
that limits their robustness.

For the Helmholtz EOS, the 2D and 3D NR solvers recover most of the 
parameter space with less than 10 iterations. Both solvers, however, need 
$\geq\!30$ iterations for very low temperatures. The 2D NR solver 
additionally needs $\gtrsim\!15$ iterations in the upper left 
corner. The extent of this region is similar to the 
results of the 2D NR solver for the LS220 EOS. The 1D Brent, 2D NR Noble, 
and 1D Newman methods recover the correct solution with fewer than 10--12 
iterations in the entire parameter space.

Based on the number of iterations to converge, the 1D Brent method would be 
one obvious first choice because it recovers the entire parameter space for all 
EOSs well with $\leq\!10$ iterations.

In the bottom panels of Fig.~\ref{fig:W1-2_prat--3}, we show the number of 
EOS calls for each recovery scheme and the same three EOSs as in the top 
panels. While there is no drastic 
variation in the number of iterations to convergence for the different 
schemes, the number of EOS calls required by the different schemes varies dramatically. All schemes need a small number ($<10$) of EOS 
calls to converge for the ideal-gas EOS. For the LS220 
EOS, the 2D and 3D NR schemes need fewer than $\sim\!25$ EOS calls to converge in 
the entire parameter space.
The 1D Brent, 2D NR Noble, and 1D 
Newman schemes, however, require typically an order of magnitude more EOS 
calls to converge throughout the entire parameter space (in the region of 
500--1000 EOS calls). The color bar is adjusted to enhance the dynamic 
range for the more efficient schemes. The average number of EOS calls of 
all test points is summarized in Table~\ref{tab:summary_rhoT}. These large 
numbers of EOS calls are due to the fact that the 1D Brent, 2D NR Noble, 
and 1D Newman schemes each require an additional inversion in the EOS 
table per solver iteration. Each of these inversions, in turn, typically 
requires tens of steps to converge and a few EOS calls per step. The results for the Helmholtz EOS are similar to those for LS220, although the absolute numbers of 
required EOS calls are lower for the 1D Brent, 2D NR Noble, and 1D Newman schemes.

\begin{table*}[tb]
\caption{Comparison of different recovery schemes for the parameters in Figs.~\ref{fig:W1-2_prat--3} and \ref{fig:W1-2_prat--3_acc}.}
\centering
\begin{tabular}{lccccc}
\hline
\hline
 EOS & 2D NR & 3D NR & 1D Brent & 2D NR Noble & Newman \\
 \hline
\multicolumn{2}{l}{Iterations:} &&&&\\
IG & 5.5 & 5.4 & 4.8 & 4.5 & 4.9 \\
LS220 & 11.5 &  9.1  & 6.9 &  6.1 &  6.1 \\
Helm & 25.0 &  16.2 &  5.7 &  6.3 &  5.9 \\
\hline
\multicolumn{2}{l}{\#EOS calls:} &&&&\\
IG & 6.5 & 7.4 & 6.8 & 6.5 & 5.9\\
LS220 & 13.0 & 11.3 & 836 & 758 & 331\\
Helm & 28.2 & 18.2 & 111 & 83.4 & 57.0\\
\hline
\multicolumn{2}{l}{Accuracy:} &&&&\\
IG & $3.4\times 10^{-12}$  & $7.5\times 10^{-13}$ &  $2.5\times 10^{-11}$ &  $8.9\times 10^{-13}$ &  $4.4\times 10^{-12}$\\
LS220 & $7.1\times 10^{-12}$ &  $1.3\times 10^{-13}$ & $1.0\times 10^{-11}$ &  $1.7\times 10^{-12}$ & $6.1\times 10^{-13}$\\
Helm & $5.9\times 10^{-13}$ & $1.7\times 10^{-14}$ & $2.3\times 10^{-12}$ &  $3.7\times 10^{-12}$ & $3.9\times 10^{-13}$\\
\hline
\end{tabular}
\tablecomments{The table compares the mean number of iterations, mean number of EOS calls, and the resulting mean accuracy averaged in log space. Here $W=2$, $p_\mathrm{mag}/p= 10^{-3}$, $Y_e=0.1$.}
\label{tab:summary_rhoT}
\end{table*}

Fig.~\ref{fig:W1-2_prat--3_acc} shows the relative error between original 
and recovered primitive variables after the schemes have converged for the 
five different schemes and three EOSs presented in Fig.~\ref{fig:W1-2_prat--3}. The NR schemes generally perform better in this metric 
because one or two additional iterations of the solver improve the accuracy by 
orders of magnitude.

Table~\ref{tab:summary_rhoT} summarizes the different metrics (number of 
iterations, number of EOS calls, and relative accuracy) averaged over all
points in the test parameter space for the five recovery schemes and three 
EOSs tested here. The averaged values well capture the trends evident from 
Figs.~\ref{fig:W1-2_prat--3} and \ref{fig:W1-2_prat--3_acc}. 
Judged by the number of iterations, the 1D Brent, 2D NR Noble and Newman 
methods are the more efficient schemes. However, the opposite is true for 
the number of EOS calls, in which case these three schemes need up to one 
or two orders of magnitude more EOS calls on average than the 2D and 3D NR 
schemes. For the Helmholtz EOS the situation is not as dire as for LS220, 
and the number of EOS calls increases by a factor of a few relative to the 2D and 3D NR schemes with the 1D Brent, 2D NR Noble, and Newman methods. The NR and Newman 
schemes reach a higher overall relative accuracy than the 1D Brent method.

\begin{table*}[tb]
\caption{Comparison of different recovery schemes for the parameters in Fig.~\ref{fig:RHO-1e11_T-5_n_iter}.}
\centering
\begin{tabular}{lccccc}
\hline
\hline
 EOS & 2D NR & 3D NR & 1D Brent & 2D NR Noble & Newman \\
 \hline
\multicolumn{2}{l}{Iterations:} &&&&\\
IG & 6.8 & 4.1 & 7.3 & 4.6 & 5.0\\
LS220 & 7.5 & 10.3 &  7.1  &  4.3 &  7.1\\
Helm & 9.8 & 13.5 &  7.4  & 4.3  & 7.0\\
\hline
\multicolumn{2}{l}{\#EOS calls:} &&&&\\
IG & 12.1 &  8.7 &  9.3  & 6.6 &  6.0\\
LS220 & 8.5 &  17.3 & 877 & 329 & 398\\
Helm & 16.3 & 52.2 & 86.2 & 39.7 & 35.0\\
\hline
\multicolumn{2}{l}{Accuracy:} &&&&\\
IG & $1.7\times 10^{-13}$ & $4.7\times 10^{-14}$ &  $1.4\times 10^{-11}$ &  $2.6\times 10^{-13}$  & $5.3\times 10^{-13}$\\
LS220 & $3.3\times 10^{-11}$ &  $3.6\times 10^{-11}$ &  $2.9\times 10^{-11}$ & $1.4\times 10^{-11}$ &  $2.0\times 10^{-12}$\\
Helm & $3.3\times 10^{-13}$ &  $3.6\times 10^{-13}$ & $2.2\times 10^{-11}$ &  $5.8\times 10^{-13}$ & $5.9\times 10^{-13}$\\
\hline
\multicolumn{2}{l}{\% recovered:} &&&&\\
IG & 56.0 &  58.7 & 74.9 & 75.7 & 79.1\\
LS220 & 52.6 & 69.1 & 68.0 & 60.0 & 77.1\\
Helm & 56.9 & 77.3 & 75.6 & 65.0 & 77.4 \\
\hline
\label{tab:summary_Wp}
\end{tabular}
\tablecomments{The table compares the mean number of iterations, mean number of EOS calls, the resulting mean accuracy averaged in log space, and the percentage of overall recovered points. Here $\rho=10^{11}\,\mathrm{g}\,\mathrm{cm}^{-3}$, $T=5\,\mathrm{MeV}\,(5.8\times 10^{10}\,\mathrm{K})$, $Y_e=0.1$.}
\end{table*}

A comparison of the five different recovery schemes covering the
parameter space in Lorentz factor $W$ and the ratio of magnetic 
to fluid pressure $p_{\textrm{mag}}/p$ is shown in Fig.~\ref{fig:RHO-1e11_T-5_n_iter} (analogous to the top panels of Fig.~\ref{fig:W1-2_prat--3}). Here we choose a representative temperature of $T= 5\, \mathrm{MeV}$ ($5.8\times 10^{10}\,\mathrm{K}$) and a density of $\rho = 10^{11}\, \mathrm{g\,cm^{-3}}$. In general, all schemes fail for very 
strongly magnetized fluid and/or high Lorentz factors. For the ideal-gas 
EOS, the methods of 1D Brent, 2D NR Noble, and 1D Newman recover most of 
the parameter space well and with few iterations ($\leq\!10$). However, 
the 2D and 3D NR methods do not recover the correct solution beyond 
Lorentz factors of 10. For the LS220 EOS, the situation is similar to the 
ideal-gas EOS with the exception of the 3D NR method recovering points up to high Lorentz factors of $W \simeq 100$. 
For the Helmholtz EOS, similar results to those for the LS220 are obtained, with the 3D NR solver now even recovering up to higher Lorentz factors of $W\simeq 10^2-10^3$. All metrics are summarized in Tab.~\ref{tab:summary_Wp}, which shows similar trends to Tab.~\ref{tab:summary_rhoT}.

Robustness of the schemes involving an NR algorithm (2D NR, 3D NR, 2D NR Noble) is severely limited by the dependence on good initial guesses. These schemes are particularly sensitive to a good initial guess for the velocity components in the regime of high Lorentz factors. Perturbations at the level of 5\% in the primitives as considered here cause definite failure for the 2D NR scheme above Lorentz factors of $W\lesssim\!10$ for all EOSs, partial failure in this regime for the 2D NR Noble scheme and the LS220 and Helmholtz EOS, and failure for the 3D NR scheme and the ideal-gas EOS, as well as a drastically increased number of iterations in the case of the LS220 and the Helmholtz EOS (see Fig.~\ref{fig:RHO-1e11_T-5_n_iter}). Additionally, all schemes centered around an NR algorithm directly involve thermodynamic derivatives, which often must be numerically computed for tabulated EOSs. These derivatives may be thermodynamically inconsistent (i.e., the Maxwell relations may not be satisfied to machine precision) and they may be noisy, which adds additional sources of error or failure that cannot be controlled and may appear quasi-randomly.

Robustness is significantly improved in the 1D Brent and Newman schemes, which do not depend explicitly on either initial guesses or thermodynamic derivatives. In particular, the fact that there are known brackets for the 1D Brent scheme guarantees the existence of a root and convergence of the scheme. However, for three-parameter EOSs, additional inversions from specific internal energy or enthalpy to temperature are required at every iteration step of the respective main scheme, which is typically accomplished via a 1D NR algorithm. This then implicitly introduces dependences on initial guesses and thermodynamic derivatives. As we find here, inversions of specific internal energy (as in the 1D Brent method) can typically be more robust than inversions of the specific enthalpy (as in the 1D Newman scheme; see the above discussion related to Fig.~\ref{fig:W1-2_prat--3}). None of the schemes under consideration here is thus entirely independent of initial guesses. However, in contrast to the 1D Brent method, the 1D Newman scheme starts with a well-defined initial guess for the temperature (the minimum temperature for the EOS at given $\rho$ and $Y_e$, see Sec.~\ref{sec:Newman}).

\section{Discussion and Conclusion}
\label{sec:conclusion}

We performed a comparison of a number of schemes to convert conservative to primitive
variables that are used in different GRMHD codes (Sec.~\ref{sec:recovery_schemes}), employing analytic, partially tabulated, and fully tabulated microphysical EOSs. We have analyzed these methods using a well-defined uniform stand-alone testbed (Sec.~\ref{sec:comparison_setup}), covering 2D parameter spaces in either density and temperature or Lorentz factor and ratio of magnetic to fluid pressure typically encountered in astrophysical simulations. The schemes are assessed in terms of the three criteria: (i) speed (according to the number of iterations and number of EOS calls required for convergence), (ii) accuracy, and (iii) robustness (Sec.~\ref{sec:comparison_setup}).

While for the ideal-gas EOS there is not too much variation among the 
different schemes in terms of speed and accuracy (typically $\leq 10$ 
iterations and EOS calls), there are significant differences for the 
microphysical EOSs (LS220 and Helmholtz EOSs; see Tables~\ref{tab:summary_rhoT} and \ref{tab:summary_Wp} for a summary). The 2D and 3D 
NR schemes require 10--20 iterations more to converge in some parts of 
the parameter space for microphysical EOSs relative to other schemes, but 
require tens to hundreds of EOS calls fewer and are thus overall much 
faster. The additional EOS calls in the 1D Brent, 2D NR Noble, and 1D 
Newman schemes arise because of an additional temperature inversion at 
every solver iteration that requires roughly 10--100 EOS calls by itself 
(depending on the inversion scheme and parameters). For tabulated EOSs 
such as the LS220 and Helmholtz EOSs, the computational cost of a 
single EOS call can be significant. Therefore, employing the most efficient scheme (3D NR) can yield a speed-up by a factor of at least a few to 10, potentially up to a factor of 100 compared to one of the least efficient schemes. Since the cost of the recovery of primitive variables in an evolution code can be up to $\sim\!40\%$ of the total simulation cost, this can have significant performance impacts on the evolution code, and may determine whether or not a certain simulation is computationally feasible.

The 3D NR scheme clearly emerges as the fastest and most accurate one; however, it is not the most robust scheme. All schemes based on an NR algorithm (2D NR, 3D NR, 2D NR Noble) require good initial guesses and thus do not guarantee convergence. In the case of the 2D NR Noble scheme this can limit the ability to recover at high Lorentz factors (see Fig.~\ref{fig:RHO-1e11_T-5_n_iter}). When used in an evolution code, these schemes thus depend on the previous time step, which represents a severe limitation that can lead to failure, because data from the previous time step can be corrupted due to numerical error. Furthermore, these schemes explicitly depend on thermodynamic derivatives, which often need to be computed numerically and can thus be noisy and thermodynamically inconsistent. This may lead to failure. The 1D Brent and 1D Newman schemes, however, are both guaranteed to converge and do not explicitly depend on initial guesses and thermodynamic derivatives. However, for microphysical EOSs, additional inversions involving the EOS are required at every solver step, which implicitly introduces dependence on initial guesses (and thermodynamic derivatives, depending on the inversion method). Only the 1D Newman method starts from an independently defined initial guess and is thus entirely independent of the previous time step when used in an evolution code. Furthermore, this additional inversion step may fail and thus limit the robustness of the main scheme. Enthalpy inversions as in the 1D Newman scheme tend to be less robust and more prone to failure (typically at the level of a few per cent in the density--temperature parameter space) than inversions of the specific internal energy as in the 1D Brent scheme.

In conclusion, the optimal recovery scheme for a GRMHD evolution code depends on the EOS. For an analytic EOS, we suggest to employ a more robust scheme such as the 1D Brent or 1D Newman scheme, with the 1D Newman scheme being preferred because it tends to achieve higher accuracy. For microphysical (three-parameter) EOSs, we envision a combination of different schemes. In a first attempt, we suggest to employ the 3D NR scheme, which is both the most efficient and the most accurate scheme among those tested here. For points that fail in this initial pass, one can then switch to a more robust scheme (1D Brent, 1D Newman) in a second step and exploit the greater robustness in the more extreme regions of the parameter
space.

An implementation of the recovery schemes discussed in this paper is available as open-source software in \citet{Siegel2018con2primcode}\footnote{Codebase: \url{https://doi.org/10.5281/zenodo.1213306}}. A separate development version is available at \url{https://bitbucket.org/dsiegel/grmhd_con2prim}. This code package can be used in GRMHD evolution codes or as a stand-alone framework to add new EOSs and recovery schemes for systematic comparison with existing ones. We value bug reports and code contributions.


\acknowledgments

We thank F.~Foucart, W.~Kastaun, B.~D.~Metzger, and C.~D. Ott for valuable discussions. Resources supporting this work were provided by the NASA High-End Computing (HEC) Program through the NASA Advanced Supercomputing (NAS) Division at Ames Research Center. D.M.S. and P.M. acknowledge support for this work by the National Aeronautics and Space Administration through Einstein Postdoctoral Fellowship Award Numbers PF6-170159 and PF5-160140 issued by the Chandra X-ray Observatory Center, which is operated by the Smithsonian Astrophysical Observatory for and on behalf of the National Aeronautics and Space Administration under contract NAS8-03060. D.M.S. acknowledges support from NASA ATP grant NNX16AB30G.
 
\software{Helmholtz EOS (\citealt{Timmes1999,Timmes2000}; \url{http://cococubed.asu.edu/code_pages/eos.shtml}), Lattimer--Swesty EOS (\citealt{Lattimer1991}; \url{http://www.astro.sunysb.edu/dswesty/lseos.html}; \url{https://stellarcollapse.org/equationofstate})}

\newpage
\bibliographystyle{aasjournal}
\bibliography{references}

\begin{thebibliography}{}
\expandafter\ifx\csname natexlab\endcsname\relax\def\natexlab#1{#1}\fi
\providecommand{\url}[1]{\href{#1}{#1}}

\bibitem[{{Aasi} {et~al.}(2014){Aasi}, {Abbott}, {Abbott}, {Abbott},
  {Abernathy}, {Acernese}, {Ackley}, {Adams}, {Adams}, {Addesso}, \&
  et~al.}]{Aasi2014}
{Aasi}, J., {Abbott}, B.~P., {Abbott}, R., {et~al.} 2014, \prl, 113, 011102

\bibitem[{{Anderson} {et~al.}(2008){Anderson}, {Hirschmann}, {Lehner},
  {Liebling}, {Motl}, {Neilsen}, {Palenzuela}, \& {Tohline}}]{Anderson2008}
{Anderson}, M., {Hirschmann}, E.~W., {Lehner}, L., {et~al.} 2008, \prl, 100,
  191101

\bibitem[{{Ant{\'o}n} {et~al.}(2006){Ant{\'o}n}, {Zanotti}, {Miralles},
  {Mart{\'{\i}}}, {Ib{\'a}{\~n}ez}, {Font}, \& {Pons}}]{Anton2006}
{Ant{\'o}n}, L., {Zanotti}, O., {Miralles}, J.~A., {et~al.} 2006, \apj, 637,
  296

\bibitem[{{Arnowitt} {et~al.}(2008){Arnowitt}, {Deser}, \&
  {Misner}}]{Arnowitt2008}
{Arnowitt}, R., {Deser}, S., \& {Misner}, C.~W. 2008, Gen. Rel. Grav., 40, 1997

\bibitem[{{Cerd{\'a}-Dur{\'a}n} {et~al.}(2008){Cerd{\'a}-Dur{\'a}n}, {Font},
  {Ant{\'o}n}, \& {M{\"u}ller}}]{Cerda-Duran2008}
{Cerd{\'a}-Dur{\'a}n}, P., {Font}, J.~A., {Ant{\'o}n}, L., \& {M{\"u}ller}, E.
  2008, \aap, 492, 937

\bibitem[{{Del Zanna} {et~al.}(2007){Del Zanna}, {Zanotti}, {Bucciantini}, \&
  {Londrillo}}]{DelZanna2007}
{Del Zanna}, L., {Zanotti}, O., {Bucciantini}, N., \& {Londrillo}, P. 2007,
  \aap, 473, 11

\bibitem[{Duez {et~al.}(2005)Duez, Liu, Shapiro, \& Stephens}]{Duez05MHD0}
Duez, M.~D., Liu, Y.~T., Shapiro, S.~L., \& Stephens, B.~C. 2005, \prd, 72,
  024028, astro-ph/0503420

\bibitem[{{Etienne} {et~al.}(2015){Etienne}, {Paschalidis}, {Haas},
  {M{\"o}sta}, \& {Shapiro}}]{Etienne2015a}
{Etienne}, Z.~B., {Paschalidis}, V., {Haas}, R., {M{\"o}sta}, P., \& {Shapiro},
  S.~L. 2015, \cqg, 32, 175009

\bibitem[{{Gammie} {et~al.}(2003){Gammie}, {McKinney}, \&
  {T{\'o}th}}]{Gammie2003}
{Gammie}, C.~F., {McKinney}, J.~C., \& {T{\'o}th}, G. 2003, \apj, 589, 444

\bibitem[{{Giacomazzo} \& {Rezzolla}(2007)}]{Giacomazzo2007}
{Giacomazzo}, B., \& {Rezzolla}, L. 2007, \cqg, 24, 235

\bibitem[{{Kiuchi} {et~al.}(2012){Kiuchi}, {Kyutoku}, \&
  {Shibata}}]{Kiuchi2012b}
{Kiuchi}, K., {Kyutoku}, K., \& {Shibata}, M. 2012, \prd, 86, 064008

\bibitem[{{Komissarov}(2005)}]{Komissarov2005}
{Komissarov}, S.~S. 2005, \mnras, 359, 801

\bibitem[{Lattimer \& Swesty(1991)}]{Lattimer1991}
Lattimer, J.~M., \& Swesty, F.~D. 1991, Nucl. Phys. A, 535, 331

\bibitem[{Lichnerowicz(1944)}]{Lichnerowicz44}
Lichnerowicz, A. 1944, J. Math. Pures et Appl., 23, 37

\bibitem[{{M{\"o}sta} {et~al.}(2015){M{\"o}sta}, {Ott}, {Radice}, {Roberts},
  {Schnetter}, \& {Haas}}]{Moesta2015}
{M{\"o}sta}, P., {Ott}, C.~D., {Radice}, D., {et~al.} 2015, \nat, 528, 376

\bibitem[{{M{\"o}sta} {et~al.}(2014{\natexlab{a}}){M{\"o}sta}, {Mundim},
  {Faber}, {Haas}, {Noble}, {Bode}, {L{\"o}ffler}, {Ott}, {Reisswig}, \&
  {Schnetter}}]{Moesta2014a}
{M{\"o}sta}, P., {Mundim}, B.~C., {Faber}, J.~A., {et~al.} 2014{\natexlab{a}},
  \cqg, 31, 015005

\bibitem[{{M{\"o}sta} {et~al.}(2014{\natexlab{b}}){M{\"o}sta}, {Richers},
  {Ott}, {Haas}, {Piro}, {Boydstun}, {Abdikamalov}, {Reisswig}, \&
  {Schnetter}}]{Moesta2014b}
{M{\"o}sta}, P., {Richers}, S., {Ott}, C.~D., {et~al.} 2014{\natexlab{b}},
  \apjl, 785, L29

\bibitem[{{Muhlberger} {et~al.}(2014){Muhlberger}, {Nouri}, {Duez}, {Foucart},
  {Kidder}, {Ott}, {Scheel}, {Szil{\'a}gyi}, \& {Teukolsky}}]{Muhlberger2014}
{Muhlberger}, C.~D., {Nouri}, F.~H., {Duez}, M.~D., {et~al.} 2014, \prd, 90,
  104014

\bibitem[{{Neilsen} {et~al.}(2014){Neilsen}, {Liebling}, {Anderson}, {Lehner},
  {O'Connor}, \& {Palenzuela}}]{Neilsen2014}
{Neilsen}, D., {Liebling}, S.~L., {Anderson}, M., {et~al.} 2014, \prd, 89,
  104029

\bibitem[{Newman \& Hamlin(2014)}]{Newman2014}
Newman, W.~I., \& Hamlin, N.~D. 2014, SIAM J. Sci. Comput., 36

\bibitem[{{Noble} {et~al.}(2006){Noble}, {Gammie}, {McKinney}, \& {Del
  Zanna}}]{Noble2006}
{Noble}, S.~C., {Gammie}, C.~F., {McKinney}, J.~C., \& {Del Zanna}, L. 2006,
  \apj, 641, 626

\bibitem[{{Palenzuela} {et~al.}(2015){Palenzuela}, {Liebling}, {Neilsen},
  {Lehner}, {Caballero}, {O'Connor}, \& {Anderson}}]{Palenzuela2015}
{Palenzuela}, C., {Liebling}, S.~L., {Neilsen}, D., {et~al.} 2015, \prd, 92,
  044045

\bibitem[{{Siegel} \& {Metzger}(2017)}]{Siegel2017a}
{Siegel}, D.~M., \& {Metzger}, B.~D. 2017, \prl, 119, 231102

\bibitem[{{Siegel} \& {Metzger}(2018)}]{Siegel2018a}
---. 2018, \apj, 858, 52

\bibitem[{{Siegel} \& {M{\"o}sta}(2018)}]{Siegel2018con2primcode}
{Siegel}, D.~M., \& {M{\"o}sta}, P. 2018, {GRMHD\_con2prim: a framework for the
  recovery of primitive variables in general-relativistic
  magnetohydrodynamics},  Zenodo, doi:10.5281/zenodo.1213306.
\newblock \url{https://doi.org/10.5281/zenodo.1213306}

\bibitem[{{Timmes} \& {Arnett}(1999)}]{Timmes1999}
{Timmes}, F.~X., \& {Arnett}, D. 1999, \apjs, 125, 277

\bibitem[{{Timmes} \& {Swesty}(2000)}]{Timmes2000}
{Timmes}, F.~X., \& {Swesty}, F.~D. 2000, \apjs, 126, 501

\bibitem[{{White} {et~al.}(2016){White}, {Stone}, \& {Gammie}}]{White2016}
{White}, C.~J., {Stone}, J.~M., \& {Gammie}, C.~F. 2016, \apjs, 225, 22

\end{thebibliography}

\end{document}